\documentclass[a4paper,fleqn]{cas-dc}

\usepackage[authoryear,longnamesfirst]{natbib}

\usepackage{booktabs}
\usepackage{multirow}
\usepackage{rotating}

\def\tsc#1{\csdef{#1}{\textsc{\lowercase{#1}}\xspace}}
\tsc{WGM}
\tsc{QE}
\tsc{EP}
\tsc{PMS}
\tsc{BEC}
\tsc{DE}

\setcitestyle{numbers,square,comma,sort&compress}

\begin{document}
\let\WriteBookmarks\relax
\def\floatpagepagefraction{1}
\def\textpagefraction{.001}

\shorttitle{Synthetic Letters for Seizure Frequency Extraction}

\shortauthors{Yujian Gan et~al.}

\title [mode = title]{Reproducible Synthetic Clinical Letters for Seizure Frequency Information Extraction}                      
\tnotemark[1]

\tnotetext[1]{This document is the results of the research
   project funded by Epilepsy Research Institute (grant ref 2209) and by an investigator-initiated research grant from Angelini Pharma.}


%
\author[1,2,3]{Yujian Gan}[type=editor,
                        auid=000,bioid=1,
                        orcid=0009-0002-3188-5924]






\affiliation[1]{organization={Department of Basic and Clinical Neuroscience, Institute of Psychiatry Psychology and Neuroscience, King’s College London},
    addressline={16 De Crespigny Park}, 
    city={London},
    postcode={SE5 8AB}, 
    country={United Kingdom}}

\author[1,3]{Stephen H. Barlow}

\author[1,3]{Ben Holgate}

\author[1,3]{Joe Davies}
\author[1,3]{James T. Teo}
\author[1,3]{Joel S. Winston}[orcid=0000-0002-3957-0612]

\affiliation[2]{organization={School of Electronics, Electrical Engineering and Computer Science, Queen's University Belfast},
    addressline={16A Malone Road}, 
    city={Belfast},
    postcode={BT9 5BN}, 
    state={Northern Ireland},
    country={United Kingdom}}

\author%
[1,3]
{Mark P. Richardson}
\cormark[1]
\ead{mark.richardson@kcl.ac.uk}

\affiliation[3]{organization={Centre for Epilepsy, King’s College Hospital NHS Foundation Trust},
    addressline={Denmark Hill}, 
    city={London},
    postcode={SE5 9RS}, 
    country={United Kingdom}}

\cortext[cor1]{Corresponding author}



\begin{abstract}\ \ \ \ \ \ \ \textbf{\bf Objective:}
Seizure-frequency information is critical for epilepsy research and clinical decision-making, yet it is often documented in highly variable free-text neurology clinic letters that are difficult to share and annotate at scale. We aimed to develop a reproducible and privacy-preserving framework for seizure-frequency information extraction by training models using fully synthetic, task-faithful clinical letters.

\textbf{Methods:}
We designed a structured label scheme that captures common linguistic patterns of seizure burden, including explicit rates, numeric ranges, cluster-based descriptions, seizure-free durations, unknown frequency, and explicit no-seizure statements. A high-capacity teacher language model was used to generate NHS-style synthetic epilepsy follow-up letters paired with normalized labels, step-by-step rationales, and evidence spans. We fine-tuned multiple open-weight large language models (4--14B parameters) to extract seizure frequency from full letters using either direct numeric outputs or structured labels, optionally augmented with evidence-grounded explanations. Performance was evaluated on a clinician double-checked held-out set of real clinic letters using both fine-grained and pragmatic frequency groupings.

\textbf{Results:}
Models trained purely on synthetic letters generalized to real-world letters, and structured label targets consistently outperformed direct numeric regression. With 15{,}000 synthetic training letters, representative models achieved micro-F1 up to 0.788 (fine-grained) and 0.847 (pragmatic) on the real test set, while a medically oriented 4B model reached 0.787 and 0.858, respectively. Evidence-grounded outputs supported rapid clinical verification and facilitated error analysis.

\textbf{Conclusion:}
Task-specific synthetic clinical letters with structured, evidence-grounded supervision enable reproducible development of seizure-frequency extractors without distributing sensitive patient text. The proposed framework is directly applicable to other temporally complex clinical information extraction problems where privacy constraints limit access to real narratives.
\end{abstract}


\begin{highlights}
\item Propose a reproducible synthetic-letter framework for seizure-frequency extraction.
\item Design a structured label scheme covering rates, ranges, clusters, and seizure-free durations.
\item Models trained on fully synthetic letters generalize to real clinic letters with strong F1 scores.
\end{highlights}


\maketitle

\section{INTRODUCTION}
\subsection{Background}

Epilepsy is a common neurological disorder and a major contributor to global disability and premature mortality, with substantial variation in burden across regions and age groups~\citep{THIJS2019689-y2,2025e203-y1}. 
In both routine care and research, seizure frequency is one of the most frequently used indicators of disease control and treatment response, and it is often discussed repeatedly across longitudinal follow-up. 
However, seizure counting is inherently difficult: many seizures are unwitnessed or impair awareness, and real-world documentation can be incomplete or inconsistent~\citep{https://doi.org/10.1111/epi.13727}. 
Studies of long-term seizure diary use in clinical research settings also report substantial heterogeneity in adherence and missing data, which can bias downstream analyses when seizure frequency is used as a primary endpoint~\citep{MILLER2024107379-y3}.

In modern health systems, seizure frequency information is typically embedded in unstructured clinical narratives, such as neurology clinic letters. 
These documents contain the richest contextual detail (interval history, medication changes, seizure semiology, safety counselling), but the seizure-frequency signal is expressed in highly diverse linguistic forms, including explicit rates (“2 per week”), numeric ranges (“4–5 per month”), vague quantifiers (“multiple”, “occasional”), seizure-free durations (“seizure free for 6 months”), and cluster-based patterns (“clusters twice monthly, ~3 per cluster”). In addition, frequency statements are often temporally anchored to implicit reference points (e.g., “since I last saw the patient”), requiring contextual information beyond the sentence itself for correct interpretation.
As a result, manual extraction is labor-intensive and difficult to scale, motivating automated information extraction from clinical text.

Prior work has demonstrated that seizure outcomes can be extracted from electronic health records using traditional NLP pipelines and machine-learning approaches, including systems targeting seizure type and frequency from free text~\citep{10.1093/jamia/ocac018-y4,DECKER202248-y5}. 
More recently, large language models (LLMs) have been explored for extracting epilepsy-related variables from unstructured letters, and fine-tuning strategies have been proposed to improve seizure-frequency extraction performance~\citep{Fang2025Epilepsia,holgate-etal-2025-fine,Abeysinghe2025}.
Nonetheless, extracting seizure frequency robustly remains challenging because it requires temporal normalization, reasoning over implicit time windows, and robustness to missing data and ambiguous expressions, which are common in real-world clinical correspondence.

A key bottleneck is that high-performing clinical NLP systems typically require large volumes of representative training data and careful evaluation. The community has benefited from shared-task corpora (e.g., i2b2-style challenges) and large critical-care datasets that include clinical notes under controlled access, such as MIMIC-III and MIMIC-IV~\citep{Uzuner2010Extracting-y9,10.1136/amiajnl-2011-000203-y10,Johnson2016-y11,Johnson2023-y12}. 
Yet broad reuse and open release of clinical text are constrained by privacy and governance requirements. De-identification methods have progressed substantially, including both early rule-based systems and modern neural approaches, but they are not perfect and can leave residual risk~\citep{Neamatullah2008-y13,STUBBS2015S20-y16,10.1093/jamia/ocx064-y15,FAUSTINI2026106225-y14}. 
At the same time, modern machine learning introduces additional concerns: models can leak training data through memorization, and attacks such as membership inference have demonstrated privacy vulnerabilities even when direct identifiers are removed. These constraints are especially salient for LLM development, where training scale and model capacity can amplify memorization and leakage risks~\citep{10.1145/2976749.2978318-y19,7958568-y18,carlini2021extractingtrainingdatalarge-y17}. 

To address these limitations, synthetic clinical text has emerged as a complementary strategy to enable reproducible research without distributing real patient letters. 
Work in this area has shown that neural language models can generate synthetic clinical notes with measurable utility for downstream NLP tasks and can be designed to support safer sharing than raw text~\citep{melamud-shivade-2019-towards-y20,10.1093/jamia/ocab112-y21}. 
More ambitiously, synthetic notes have been used to train clinical LLMs that transfer to real-note benchmarks, suggesting synthetic corpora can substitute for real notes in at least some settings~\citep{kweon2024publiclyshareableclinicallarge-y22}.
In parallel, recent pipelines focus on generating, validating, and correcting synthetic clinical documents (such as discharge summaries), and on leveraging medical structure/knowledge to improve fidelity~\citep{Kumichev_2024-y24,GRAZHDANSKI2025104906-y23}.
However, for a task as sensitive to phrasing and temporal logic as seizure-frequency extraction, synthetic data must be task-faithful (reflecting real documentation patterns) and label-faithful (providing reliable supervision aligned with clinically meaningful normalization).

\subsection{Significance}

This paper targets a practical and clinically meaningful goal: reliable seizure-frequency information extraction from narrative clinic letters, while reducing dependence on real patient text for model development and sharing. Building on advances in clinical NLP and LLM-based extraction~\citep{Fang2025Epilepsia,holgate-etal-2025-fine,Abeysinghe2025}, the work is significant in three main ways.

First, we propose a synthetic-data framework and a task-specific annotation scheme tailored to seizure-frequency extraction. Existing work on synthetic clinical notes primarily emphasizes general-purpose note realism or transfer to broad benchmarks~\citep{melamud-shivade-2019-towards-y20,10.1093/jamia/ocab112-y21,kweon2024publiclyshareableclinicallarge-y22,Kumichev_2024-y24,GRAZHDANSKI2025104906-y23}. 
In contrast, seizure-frequency extraction requires precise temporal interpretation and robust handling of clinically prevalent phenomena, including numeric ranges, cluster-based patterns, seizure-free intervals, and missing or ambiguous descriptions. By explicitly encoding these patterns in a structured annotation scheme, our framework provides more robust and clinically aligned supervision than generic synthetic note generation, directly addressing the gap between existing synthetic data approaches and the requirements of epilepsy outcome modelling.

Second, the approach emphasizes reproducible training and privacy-preserving deployment through fully synthetic supervision. We show that seizure-frequency extraction models can be trained entirely on synthetic clinic letters and achieve performance on held-out real clinic letters that matches or exceeds models trained solely on real data. By eliminating dependence on sensitive longitudinal correspondence, this strategy avoids both privacy risks and the irreproducibility inherent in institution-specific datasets. Importantly, the observed gains in generalisation and robustness stem from our structured label scheme and reasoning supervision, enabling synthetic data to serve as a richer and more consistent training signal than prior approaches~\citep{holgate-etal-2025-fine} based on real clinic letters. This combination of reproducibility, privacy, and performance is particularly critical for seizure-frequency extraction, where clinically informative text is rarely shareable across sites.

Third, the framework leverages and operationalizes reasoning- and evidence-aware supervision. Chain-of-thought prompting has been shown to improve multi-step reasoning in LLMs~\citep{wei2022chain,miao2024cotnephrology}, and reasoning behaviors can be transferred into smaller models via distillation and supervised fine-tuning~\citep{shridhar2023distilling}. 
For clinical information extraction, the ability to provide grounded explanations (e.g., evidence spans tied to the source text) supports auditability and human verification, aligning with broader NLP research on rationales and evidence-based evaluation~\citep{lei2016rationalizingneuralpredictions,deyoung-etal-2020-eraser}. 
In the context of epilepsy, where downstream clinical and research decisions may depend on seizure-frequency estimates, such transparency can be as important as raw accuracy.

Overall, the work contributes a concrete pathway for developing seizure-frequency extraction systems that are scalable, reproducible, and better aligned with privacy and deployment realities in healthcare, while also serving as a template for extending synthetic, evidence-grounded supervision to other clinically important extraction problems.
\section{Methods}

\subsection{Ethical approval}
This project operated under UK Health Research Authority (HRA) London South East Research Ethics Committee approval (reference 18/LO/2048 and renewed 24/LO/0057) granted to the King's Electronic Records Research Interface (KERRI) with data research opt-out. This study was approved by the KERRI committee at King's College Hospital (KCH) for purposes of evaluating NLP for Epilepsy (approved February 2021) with local institutional oversight.

\subsection{Large language models}
Given the sensitivity of real patient data in this study and the requirement for strict compliance with UK data protection standards, ensuring privacy and secure deployment is essential. For this reason, we focused on open-weight LLMs that can be run entirely within a hospital’s internal system, without the need for internet connectivity. This aligns with the practical constraints of typical NHS computational environments, where GPU resources are often limited. To balance computational feasibility with strong modelling capacity, we trained a set of relatively small to medium-scale open-source models, including Qwen2.5-7B-Instruct, Qwen2.5-14B-Instruct, Gemma-3-4B-it, MedGemma-4B-it, Lingshu-7B, Meta-Llama-3.1-8B-Instruct, and Ministral-8B-Instruct-2410. Model sizes in the 4–14 billion parameter range offer a practical trade-off: they are large enough to capture complex temporal and clinical patterns relevant for epilepsy care, yet small enough to be trained or fine-tuned with moderate GPU resources and subsequently integrated into routine clinical workflows.

A key difference between our study and previous work~\citep{holgate-etal-2025-fine} is our incorporation of GPT-5 as a high-performance teacher model. 
In this context, a teacher model refers to a large, high-capacity language model used to provide supervision signals for training or guiding smaller models, a paradigm commonly known as knowledge distillation. Such models are not deployed directly; instead, they leverage their stronger reasoning and generative capabilities to produce high-quality synthetic data or annotations that support the training and domain adaptation of more lightweight models suitable for real-world deployment.
Following this paradigm, GPT-5 was used to generate synthetic UK NHS neurology clinic letters designed to closely mimic the structure, language, and documentation patterns of real epilepsy follow-up communications~\footnote{This project respects OpenAI’s Terms of Use. This study is conducted for academic, non-commercial research purposes. We do not release distilled model weights and do not position the distilled model as a deployable substitute for OpenAI services; results are reported for methodological analysis.}.
We prepared 10 anonymized base letters together with varying descriptions of seizure frequency. By systematically combining these base templates with different seizure-frequency scenarios and prompt specifications, GPT-5 was guided to generate the final synthetic clinic letters.
Beyond generating the letters themselves, GPT-5 also produced detailed analytic outputs, including seizure-frequency reasoning, final labels, and explicit justifications highlighting which specific parts of the letter support each decision. 
These multi-layered synthetic datasets enabled several training strategies, including training exclusively on synthetic letters, training on analytic explanations, and training on combinations of both. 
This approach provides the dual benefit of avoiding the use or exposure of any proprietary or institutionally held patient data, while allowing the open-weight models to acquire richer epilepsy-specific reasoning capabilities through high-quality synthetic supervision.

\subsection{Data source}
In this study, we accessed over 25K electronic health records (EHRs) from King’s College Hospital (KCH) in London, including demographics and neurology clinic letters, spanning from January 2013 to September 2023. As our focus was on adult epilepsy care, we included only EHRs of patients aged 18 years or older with a diagnosis of epilepsy, in order to avoid the additional biases and complexities introduced by paediatric populations. Access to KCH’s internal data management platform enabled efficient filtering of the required records according to these criteria.

We restricted our analysis to EHRs generated during attendances at secondary or tertiary care epilepsy ambulatory clinics or outpatient clinics. In the UK, these encounters are typically documented as an electronic note written by the clinician (usually a neurologist or epilepsy nurse specialist), which is then formatted as a letter addressed to the patient and copied to the primary care physician involved in the patient’s care. Such letters usually summarise the patient’s history, examination findings, relevant investigations, current antiseizure treatment, interval progress, and management plans. Additional notes may be present, including communications with other health care professionals, the patient, or their carers. In this work, we collectively refer to these documents as “clinic letters”.

From the full pool of eligible records, we selected 1,781 clinic letters for detailed annotation. Of these, 300 letters that underwent a double-checking procedure to ensure label quality were reserved as a held-out test set, while the remaining 1,481 letters were used for training.
Each letter corresponded to a single visit of an individual patient. 
In contrast to previous studies~\citep{Fang2025Epilepsia} that have focused on extracting relatively well-defined clinical attributes, such as epilepsy type, seizure classification or medication lists, our task centred on seizure frequency, a substantially more complex form of information to recover from unstructured text.

In addition to the real-world KCH data, we constructed a large synthetic corpus to support model training. Using GPT-5 as a teacher model, we generated 15,099 synthetic UK NHS style neurology clinic letters that mimic the structure and linguistic characteristics of genuine epilepsy follow-up correspondence, but contain no real patient information. These synthetic letters were created to reflect a wide range of seizure frequency patterns and clinical contexts, and were paired with structured outputs (e.g. seizure frequency labels and supporting evidence spans) to facilitate supervised learning. Together, the real and synthetic datasets provide a scalable yet privacy-preserving basis for training and evaluating LLMs to infer seizure frequency from unstructured clinical text.

\subsection{Chain of Thought Prompting}

Chain-of-thought (CoT) prompting encourages models to generate explicit intermediate reasoning steps before producing a final answer, and has been shown to improve performance on multi-step reasoning tasks~\citep{wei2022chain}. These reasoning traces can also be transferred from larger teacher models to smaller students through supervised fine-tuning~\citep{shridhar2023distilling}, and recent work has highlighted the value of such rationales in clinical contexts where transparency is essential~\citep{miao2024cotnephrology,wu2024medcot}.

In our study, GPT-5 was prompted not only to generate synthetic UK NHS neurology clinic letters, but also to produce a structured chain of thought for each letter. These chains of thought identify seizure-related spans, describe how temporal information is interpreted, and compute the final normalised seizure-frequency label. 

These GPT-5–generated chains of thought were then used as supervision signals when fine-tuning the open-weight models. Each training instance consisted of a clinic letter as input and a structured target output containing the reasoning steps, the final seizure-frequency label, and the specific evidence spans from the letter that support the conclusion. This enables smaller models to learn not only the final prediction but also how the teacher model grounds its reasoning in the underlying clinical text.

At inference time on real KCH clinic letters, models were prompted to “first explain, then answer.’’ For evaluation, only the final categorical and numerical seizure-frequency outputs were compared with the gold-standard labels. The explicit chains of thought remain valuable for clinical interpretability: they allow clinicians to understand why a particular label was produced, reducing the extent to which the model behaves as a black box and enabling clinicians to judge whether the prediction is supported by the evidence in the letter.

\subsection{Real clinic letter annotation} 
To construct a high-quality reference dataset for training and evaluating LLMs, we performed manual annotation of seizure frequency on the 1,791 real clinic letters obtained from KCH. These letters—each corresponding to a single outpatient neurology visit—contain free-text narrative documentation of patients’ seizure history, treatment, and clinical progress.
Because seizure frequency is often described implicitly or with considerable variation in phrasing, expert annotation was required to establish reliable ground-truth labels.

{\bf Annotation schema.}
Each clinic letter was annotated using a predefined set of seizure-frequency categories designed to capture the spectrum of temporal seizure patterns commonly encountered in epilepsy care. For each letter, annotators marked every category as present (“1”) or absent (“0”) based solely on information contained in the text.

\begin{table*}[h!]
\centering
\caption{Seizure frequency categories and performance evaluation methods.}
\label{tab:freq_real}
\renewcommand{\arraystretch}{1.15}
\setlength{\tabcolsep}{6pt}

\begin{tabular}{l c l c l}
\toprule
\multicolumn{3}{c}{\textbf{Purist Method}} & \multicolumn{2}{c}{\textbf{Pragmatic Method}} \\
\cmidrule(lr){1-3}\cmidrule(lr){4-5}
\textbf{Categories} & \textbf{Range ($x$)} & \textbf{Abbrev.} & \textbf{Range ($x$)} & \textbf{Abbrev.} \\
\midrule

$< 1$ per 6 months
& $0 < x \le 0.16$
& $<$1/6M
& \multirow{4}{*}{$0 < x \le 1.1$}
& \multirow{4}{*}{infrequent}
\\

= 1 per 6 months
& $0.16 < x \le 0.18$
& 1/6M
& &
\\

$> 1$ per 6 months, $< 1$ per month
& $0.18 < x \le 0.99$
& (1/6M,1/M)
& &
\\

= 1 per month
& $0.99 < x \le 1.1$
& 1/M
& &
\\

\hline

$> 1$ per month, $< 1$ per week
& $1.1 < x \le 3.9$
& (1/M,1/W)
& \multirow{4}{*}{$1.1 < x \le 999$}
& \multirow{4}{*}{frequent}
\\

= 1 per week
& $3.9 < x \le 4.1$
& 1/W
& &
\\

$> 1$ per week, $< 1$ per day
& $4.1 < x \le 29$
& (1/W,1/D)
& &
\\

$\ge 1$ per day
& $29 < x \le 999$
& $\ge$1/D
& &
\\

\hline

Unknown
& $x = 1000$
& UNK
& $x = 1000$
& UNK
\\

\hline

No seizure
& $x = 0$
& NS
& $x = 0$
& NS
\\

\bottomrule
\end{tabular}

\vspace{2pt}
\footnotesize{$x$ denotes seizure frequency measured as seizures per month.}
\end{table*}

{\bf Conversion to standardized numerical labels.} Following the clinicians’ categorical annotation, our data science team converted category-level labels into a unified numerical representation reflecting the number of seizures per month. Categories corresponding to less than one seizure per month were mapped to decimal values (e.g., 1 per year → 0.083 per month), while higher-frequency categories were mapped to their corresponding integer counts. 

For clarity of analysis, we consolidated the clinician-defined categories into two harmonised classes. The first class, “Unknown”, includes letters that describe seizures without providing a clear or interpretable time frame, as well as letters that contain no seizure-related information. In prior work~\citep{holgate-etal-2025-fine}, these were treated as separate categories (“Unknown frequency” and “No seizure information”). In the present study, however, we additionally introduce a distinct “No seizure” class to capture letters that explicitly document the absence of seizures (including seizure-free presentations). To maintain a comparable overall classification structure and enable fair evaluation against previous approaches, we therefore merged the earlier two categories into a single “Unknown” class.
All categories are listed in Table~\ref{tab:freq_real}. This revised categorisation allows us to separately assess model performance in cases where seizure activity is explicitly absent, while preserving comparability with prior studies, thereby supporting a more comprehensive evaluation of real-world clinical applicability.

{\bf Gold-standard test set.}
Of the annotated dataset, 300 letters underwent an additional double-review process by two senior, experienced clinicians specialising in epilepsy, in order to establish a high-quality gold-standard test set.

\begin{figure}
    \centering
    \includegraphics[width=1\linewidth]{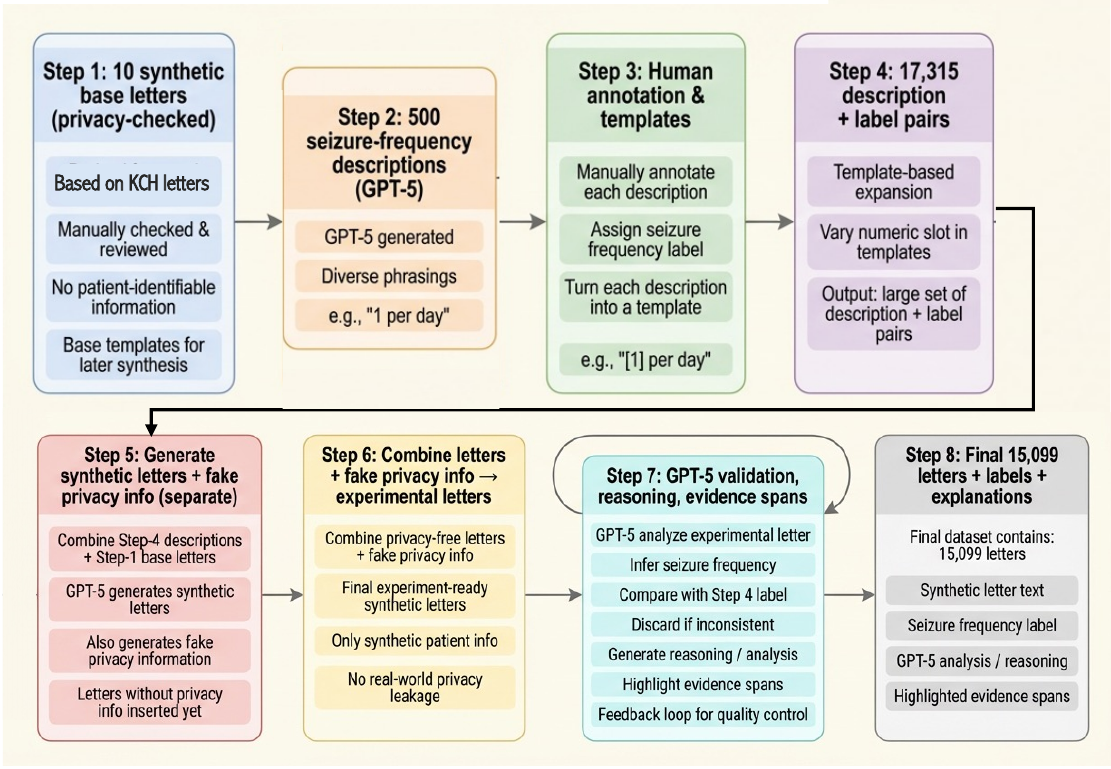}
    \caption{Synthetic Letter Generation Pipeline}
    \label{fig:Synthetic Letter Generation Pipeline}
\end{figure}

\subsection{Synthetic Data Framework} 
To obtain a supervision signal that can be publicly shared while avoiding any leakage of real patient information, we construct a fully synthetic corpus of NHS-style clinic letters. The overall framework is illustrated in Figure~\ref{fig:Synthetic Letter Generation Pipeline}. 

In Step~1, we create a set of ten synthetic ``base letters''. Each base letter is based on a real KCH clinic letter but is manually checked and edited to remove any potentially identifiable information, resulting in privacy-safe templates that preserve clinical style and structure. 
In Step~2, we use GPT-5 to generate 500 diverse short descriptions of seizure frequency, aiming to cover the range of ways clinicians might naturally express seizure burden. 
We selected GPT-5 as the synthetic data generator because it represents a high-capacity frontier model with strong long-context reasoning and advanced language generation capabilities. Beyond its modelling advantages, GPT-5 enables rapid and scalable generation of clinically realistic text. The complete synthetic corpus was produced at an approximate cost of £300.

Step~3 establishes our label scheme: human annotators read each GPT-5 description, assign the corresponding seizure-frequency label, and rewrite the text into a parametric template with a slot that will later be randomly instantiated (e.g.\ ``1 per day'' $\rightarrow$ ``[1] per day''). 
In Step~4, we systematically instantiate these templates to obtain 17{,}315 final (description, label) pairs, covering a diverse array of seizure frequencies and linguistic realisations.

Steps~5 and~6 then lift these description-label pairs to full clinic letters. 
In Step~5, GPT-5 is conditioned on one of the ten base letters and one of the synthetic descriptions to draft a new letter that (i) replaces the original seizure paragraph with the new description and (ii) generates a matching set of synthetic demographic details (name, address, identifiers, etc.). 
Crucially, these privacy-like fields are kept separate from the letter body at this stage. 
In Step~6, we perform a controlled post-processing step: the letter text contains explicit placeholders such as \texttt{@NAME@} and \texttt{@ADDRESS@}, which are then filled with the separately generated synthetic details. 
This design allows us to release letters without including our own synthetic identities, while still making it easy for other groups to recreate realistic letters by sampling their own fake demographic data.

Finally, in Step~7, each completed synthetic letter from Step~6 is sent back to GPT-5, which is asked to infer the seizure-frequency label and provide a short free-text explanation with supporting evidence spans. Only letters where the inferred label exactly matches the ground-truth label from Step~4 are kept; all others are discarded. 
Step~8 yields the final dataset of 15{,}099 synthetic letters, each paired with a normalised seizure-frequency label, a model-generated explanation, and explicit pointers to the parts of the letter that support the label.


\subsubsection{Label Scheme}
\label{sec:Label Scheme}
This subsection corresponds to Step~3 in Figure~\ref{fig:Synthetic Letter Generation Pipeline}. 
To ensure that the synthetic descriptions generated in Step~2 can be used as reliable supervision targets, we design a structured label scheme that captures the main linguistic patterns about seizure frequency. Each description is mapped to a single canonical label chosen from one of the following formats:

\begin{itemize}
    \item \texttt{unknown}
    \item \texttt{no seizure frequency reference}
    \item \texttt{seizure free for <value1|multiple> <month|year>}
    \item \texttt{<value1|multiple> per <value2|multiple> \\ <day|week|month|year>}
    \item \texttt{<value1|multiple> cluster per <value2|multiple> <day|week|month|year>, <value3|multiple> per cluster}
    \item \texttt{unknown, <value1|multiple> per cluster}
\end{itemize}

Here, \texttt{<value1>}, \texttt{<value2>} and \texttt{<value3>} represent numerical quantities that may be given explicitly (e.g.\ \texttt{1 per week}) or as ranges (e.g.\ \texttt{2 to 3 per day}). When the text specifies recurrent seizures without a clear count, we use the keyword \texttt{multiple}, such as \texttt{multiple per month}. This allows us to maintain a consistent format even when only partial quantitative information is available.

The label \texttt{unknown} is used when seizures are mentioned but no interpretable frequency can be extracted from the description. The label \texttt{no seizure frequency reference} captures cases where seizures are not mentioned at all. 
Duration-based seizure freedom is represented using the pattern \texttt{seizure free for <numeric value|multiple> <month|year>}. For example, “seizure free for 2 years” is normalised to \texttt{“seizure free for 2 year”}, and non-specific expressions such as “several months” are normalised to \texttt{“seizure free for multiple month”}.

Most narrative descriptions of seizure burden can be captured by the rate pattern \texttt{<value1|multiple>} \texttt{per} \texttt{<value2|}\\
\texttt{multiple>} \texttt{<day|week|month|year>}. For cluster-based descriptions, we use a two-level representation: \texttt{<value1|multiple> cluster per <value2|multiple> <day|week|month|year>, <value3|multiple> per cluster}. This separates the temporal spacing of clusters from the number of seizures in each cluster. 
Finally, the label \texttt{unknown, <value1|multiple> per cluster} is used when the text describes seizures occurring in clusters and an approximate within-cluster count can be inferred, but the spacing between clusters is not recoverable.

Overall, this label scheme covers the major patterns found in real clinical correspondence: explicit numeric rates, ranges, cluster-based structures, duration-based seizure freedom, and cases where frequency information is absent or fundamentally ambiguous. 
All labels can be deterministically mapped to a normalised “seizures per month” target for downstream model training and evaluation by applying a fixed set of normalisation rules. For example, non-specific quantifiers such as \texttt{multiple} are mapped to a predefined numeric value (e.g., \texttt{multiple} $\rightarrow$ 3) to enable consistent conversion into a single continuous scale.

\subsubsection{Synthetic Letters} 
This subsection corresponds to Steps~4--6 in Figure~\ref{fig:Synthetic Letter Generation Pipeline}. After defining the label scheme, we expand the annotated templates into a large collection of synthetic descriptions and then embed these descriptions into full clinic letters.

In Step~4, each template from Step~3 is instantiated with a set of allowed values, yielding 17{,}315 distinct (description, label) pairs. 
Because the templates are label-aware by construction, every generated description comes with a precise seizure-frequency label.

Steps~5 and~6 turn these description--label pairs into complete synthetic NHS letters. We first ask GPT-5 to take as input (i) one of the ten privacy-checked base letters from Step~1 and (ii) one synthetic seizure-frequency description from Step~4. GPT-5 then rewrites the letter so that the seizure-related paragraph reflects the new description, while preserving the overall clinical structure and tone of the original base letter. At the same time, GPT-5 is instructed to propose a set of fully synthetic demographic details (patient name, address, identifiers, GP details, etc.), but these are returned as a separate structured record rather than being directly embedded into the text.

To maintain strict control over privacy, the letter produced in Step~5 contains placeholder tokens such as \texttt{@NAME@} and \texttt{@ADDRESS@}, instead of concrete identifiers. 
In Step~6, a simple post-processing script replaces these placeholders with the corresponding synthetic values from the structured record, yielding final experiment-ready letters in which all apparent personal information is synthetic. 
This two-stage design has two advantages: (i) the dataset we release does not need to include any synthetic identities at all---other groups can regenerate their own fake demographics and insert them into the placeholders if required; and (ii) we do not release any synthetic identifiers at all, so even if a generated identifier happened to resemble a real one, it would never appear in the public dataset, eliminating this class of privacy risk entirely.

\subsubsection{Synthetic Chain of Thought Analysis} 
This subsection corresponds to Step~7 in Figure~\ref{fig:Synthetic Letter Generation Pipeline}. For many simple descriptions, GPT-5 can correctly infer the corresponding label directly from the letter text. 
However, we observed that even state-of-the-art LLMs such as GPT-5, even when operating in reasoning mode, can struggle with complex narratives.
To stabilise performance on these harder cases, we augment a subset of synthetic letters with chain-of-thought (CoT) style rationales.

Concretely, for a pool of challenging templates we manually design CoT exemplars that spell out the intermediate reasoning steps required to derive the correct label. During dataset construction, GPT-5 is first asked to read a synthetic letter and output both a seizure-frequency label and a free-text explanation. If the predicted label matches the ground-truth label from Step~4, we keep the model-generated explanation as-is. If the label disagrees, we then provide the model with the manual CoT exemplar and ask it to recompute the label and explanation conditioned on this guidance. 
Examples that still fail to match the ground-truth label after seeing the CoT are discarded entirely.

The result is a curated set of 15{,}099 letters for which GPT-5 not only recovers the correct label but also produces a coherent step-by-step rationale and highlights the specific parts of the letter that support the inferred seizure frequency. 
These chain-of-thought annotations can be used both as training signals for explainable models and as a diagnostic resource for analysing failure modes.

\begin{table}[t]
\centering
\caption{Multi-pass screening results.}
\label{tab:screening}
\renewcommand{\arraystretch}{1.1}
\setlength{\tabcolsep}{4pt}
\begin{tabular}{lccc}
\toprule
Method & Pass 1 & Pass 2 & Pass 3 \\
\midrule
with analysis    & 211 & 150 & 127 \\
without analysis & 250 & 190 & -- \\
\bottomrule
\end{tabular}

\vspace{2pt}
\footnotesize{Values indicate the number of discarded cases at each pass.}
\end{table}

\subsubsection{Iterative Verification and Filtering}

Table~\ref{tab:screening} summarises the number of instances that fail verification at each stage of the iterative screening process (corresponding to Step~7 in Figure~\ref{fig:Synthetic Letter Generation Pipeline}). The initial candidate pool contains 1,924 synthetic instances. Importantly, only around 30\% of these instances include an explicit \texttt{analysis} field.

In our initial setup, no explicit analysis was provided. Under this setting, nearly 10\% of instances were discarded during verification, with failures disproportionately concentrated among more complex cases. Notably, these failures persist even when using GPT-5 in reasoning mode, indicating that extracting seizure frequency from the complex clinical narratives remains challenging even for state-of-the-art models.

To address this limitation, we introduce explicit \texttt{analysis} supervision for complex instances. These analyses help GPT-5 interpret the clinical narrative and disambiguate how seizure-frequency information is expressed in the letter. As shown in Table~\ref{tab:screening}, the inclusion of analysis consistently reduces the number of discarded instances at each verification stage, allowing substantially more difficult examples to be retained in the final dataset.

\subsection{Extraction}
\label{subsec:extraction}

\cite{holgate-etal-2025-fine} found that extracting seizure frequency from clinic letters using off-the-shelf general-purpose LLMs is challenging. 
Real-world documentation exhibits substantial linguistic and temporal variability (e.g., numeric ranges, vague quantifiers and inconsistent time windows). 
In addition, under realistic computational constraints (limited model capacity and GPU resources), it is difficult to enforce stable, structured outputs; hallucinations and format drift frequently break downstream parsing and evaluation.

Subsequent work has further demonstrated the potential of large language models for seizure-frequency extraction in epilepsy reports, including multi-stage prompting pipelines and combinations of commercial and pretrained models \citep{Abeysinghe2025}. These studies highlight that LLMs can achieve strong performance when extracting structured attributes from clinical text. However, they primarily focus on task-level performance and do not systematically examine how output representation design affects trainability, decoding stability, and clinical interpretability under realistic deployment constraints.

To address this, we trained seizure-frequency extraction LLMs and designed four target output formats, each paired with a corresponding prompt, to systematically compare trade-offs between trainability, evaluability, and clinical interpretability.
Given a full clinic letter as input, the model output is constrained to one of the following four formats:

\textbf{Format 1: numeric regression-style output ($x$ per month).}
Consistent with our clinician annotation, the model outputs a single normalised numeric value $x$ representing seizures per month. 
This format is the most compact and directly supports quantitative analysis and alignment with the numeric intervals in Table~\ref{tab:freq_real}. 
However, when the letter states a \emph{range} (e.g., ``4--5 per month'') or uses \emph{vague} frequency descriptors (e.g., ``several per week''), a single point estimate inevitably discards uncertainty. In practice, the model is forced to ``pick a number'', which can reduce the interpretability of errors and complicate qualitative review.

\textbf{Format 2: four-way categorical output (pragmatic class).}
We cast extraction as a classification task and require the model to output exactly one of four categories: \texttt{frequent seizures}, \texttt{infrequent seizures}, \texttt{unknown}, or \texttt{no seizures}. 
Here, all categories correspond to the pragmatic frequency grouping in Table~\ref{tab:freq_real}. 
This format is typically stable for training and evaluation due to a small output space and high tolerance to minor generation variability, substantially reducing failures caused by formatting errors. 
The trade-off is reduced resolution: it cannot express ranges, seizure-free durations, or finer-grained frequency structure, and is therefore less informative for detailed modelling and clinical interpretation.

\textbf{Format 3: structured label output (Label Scheme).}
To better balance evaluability and expressivity, we require the model to output a label in the \emph{Label Scheme} defined in Section~\ref{sec:Label Scheme}. 
Compared with Formats 1--2, this representation naturally captures ranges, vague quantifiers, and seizure-free duration expressions while remaining strictly machine-parsable. Importantly, each label can be \emph{deterministically mapped} to a normalised ``seizures per month'' target for unified evaluation and alignment with the real-world annotation. 
Because the output structure is closer to the surface form of clinical narratives, this format can also reduce the amount of implicit reasoning required to ``force'' a numeric output.

\textbf{Format 4: CoT + label + evidence.}
To improve interpretability and auditability, we extend Format 3 with explicit reasoning and evidence grounding. 
The model outputs a structured object containing (i) a chain-of-thought style analysis, (ii) the final Label Scheme label, and (iii) supporting evidence spans (references) copied from the input letter. 
The \texttt{reference} field forces the model to explicitly anchor its decision in the source text (e.g., the sentence containing the frequency statement). This provides practical benefits for error analysis and clinical review: when a prediction is incorrect, we can distinguish reading/normalisation failures from evidence-localisation failures, and clinicians can rapidly verify whether the output is supported by the letter. The main cost is increased output length and stronger formatting constraints, which place higher demands on decoding stability and instruction-following.

\textbf{Example across formats.}
To illustrate the differences, consider a clinic letter containing the statement:
\emph{``He continues to experience four to five seizures a month on average ...''}
The corresponding outputs under each format are:
\begin{enumerate}
    \item \textbf{Format 1 (x per month):} \texttt{4}.
    \item \textbf{Format 2 (Pragmatic  class):} \texttt{frequent seizures}.
    \item \textbf{Format 3 (Label Scheme):} \texttt{4 to 5 per month}.
    \item \textbf{Format 4 (COT+label+evidence):}
    \texttt{\{"analysis": "The letter explicitly states: 'He continues to experience four to five seizures a month on average ...'. This provides a clear numeric range for current seizure frequency with no conflicting statement. Therefore the normalised label is '4 to 5 per month'.", 
    "seizure\_frequency\_number": ["4 to 5 per month", "He continues to experience four to five seizures a month on average ..."]}
\end{enumerate}

This example highlights that Formats 3 and 4 preserve the frequency \emph{range} (``4--5''), whereas Format 1 compresses it into a single-point estimate and Format 2 further abstracts it into a coarse category. In addition, Format 4 provides an explicit rationale and evidence span, facilitating clinician verification and improving trustworthiness in practice.

\subsection{Training}
\label{subsec:training}

All models were fine-tuned using the open-source\\ \texttt{LLaMA-Factory} framework, which provides a unified and reproducible training pipeline for instruction-tuned large language models. Despite its name, the framework is not affiliated with Meta and is not restricted to LLaMA-family models; it supports a wide range of open-weight transformer architectures. We selected LLaMA-Factory over custom training scripts because it offers a stable, well-tested infrastructure for configuration management, parameter-efficient fine-tuning, and distributed optimisation, thereby improving reproducibility and reducing engineering overhead.
Training was conducted for three epochs on a cluster of 8 NVIDIA V100 GPUs (32\,GB memory per GPU). To improve memory efficiency and training throughput, we employed DeepSpeed for distributed training and optimisation.

We fine-tuned seven open-weight LLMs spanning different parameter scales and design goals. These include general-purpose models (Meta-Llama-3.1-8B-Instruct, \\
Qwen2.5-7B-Instruct, Qwen2.5-14B-Instruct, Ministral-8B-Instruct-2410, and Gemma-3-4B-it) as well as two medical-oriented model (Lingshu-7B and MedGemma-4B-it). Together, these models cover a range of architectures (4B--14B parameters) and domain specialisation, allowing us to assess how model scale and medical pretraining affect seizure-frequency extraction performance under identical training and evaluation settings.

We trained models using combinations of real clinic letters and synthetic letters, depending on the extraction format. A key constraint arises from the mismatch between annotation schemes: real KCH clinic letters are annotated only with numeric seizure-frequency labels (Format~1: \emph{x per month}), whereas synthetic data are annotated using the full Label Scheme and, for some settings, include chain-of-thought explanations and evidence spans.

As a result, not all output formats can be trained on mixed real and synthetic data:

\begin{itemize}
    \item \textbf{Formats 1 (x per month) and 2 (Pragmatic class):}  
    These formats support joint training on real and synthetic data. Synthetic labels are first converted into Format~1 numeric values (seizures per month), ensuring compatibility with real letter annotations. Format~2 labels are then deterministically derived from Format~1 values using the pragmatic frequency grouping in Table~\ref{tab:freq_real}. This allows both real and synthetic letters to be combined into a single training corpus.

    \item \textbf{Formats 3 (Label Scheme) and 4 (CoT + label + evidence):}  
    These formats cannot be trained jointly with real clinic letters, as the real data do not contain Label Scheme annotations or chain-of-thought supervision. Models targeting Formats~3 and~4 are therefore trained exclusively on synthetic data, where structured labels and reasoning traces are available by construction.
\end{itemize}

\begin{table}[t]
\centering
\caption{Class distribution (support) of seizure frequency labels in Real(300), Real(150), and Synthetic(1166).}
\label{tab:real300_dist_pragmatic}
\renewcommand{\arraystretch}{1.15}
\begin{tabular}{lrrr}
\toprule
Class & Real(300) & Real(150) & Synthetic(1166)\\
\midrule
$<$1/6M       & 5   & 5   & 32  \\
= 1/6M          & 2   & 1   & 9   \\
(1/6M,1/M)      & 19  & 17  & 103   \\
= 1/M           & 6   & 6   & 52 \\
(1/M,1/W)       & 19  & 19  & 184 \\
= 1/W           & 4   & 4   & 27 \\
(1/W,1/D)       & 26  & 26  & 287 \\
$\ge$1/D        & 23  & 22  & 12  \\
UNK             & 163 & 25  & 316 \\
NS              & 33  & 25  & 144 \\
\midrule
infrequent      & 32  & 29  & 196 \\
frequent        & 72  & 71  & 510 \\
UNK             & 163 & 25  & 316 \\
NS              & 33  & 25  & 144 \\
\bottomrule
\end{tabular}
\end{table}


\subsection{Evaluation}
\label{subsec:evaluation}

We evaluate model performance by comparing predicted seizure-frequency outputs against clinician-annotated gold-standard labels under both the \emph{Purist} and \emph{Pragmatic} evaluation schemes defined in Table~\ref{tab:freq_real}.
Because the four extraction formats produce outputs at different levels of granularity, all predictions are first converted into a common categorical frequency representation before evaluation.
Classification metrics are then computed on these aligned categories.
All evaluations are performed with decoding temperature set to 0 to ensure deterministic outputs.





\subsubsection{Label distribution.}
For transparency, we report the label distribution of each evaluation dataset (Real (300), Real (150), and Synthetic (1,166)) under both the Purist and Pragmatic schemes.
Real (150) is constructed from Real (300) by selecting a subset to make category counts as even as possible across seizure-frequency classes.
The Synthetic (1,166) evaluation set was constructed via random sampling, with an additional constraint to encourage a more balanced pragmatic category distribution.
A summary of evaluation-set label distributions is provided in Table~\ref{tab:real300_dist_pragmatic}.

\subsubsection{Metrics.}
We report F1 scores using micro, macro, and weighted averaging.
Due to the pronounced class imbalance in the evaluation datasets, especially under the fine-grained Purist scheme where several categories have very low support, macro-averaged metrics can be unstable and overly influenced by rare classes.
Therefore, we include micro, macro, and weighted F1 scores in the summary tables to provide a comprehensive view, and subsequently \textbf{focus on Micro F1} as the primary metric.

\subsubsection{Uncertainty estimates.}
For each model--training-set configuration, we performed four independent fine-tuning runs with different random seeds (affecting data shuffling and optimisation).
We report the mean F1 score across runs and the standard deviation (SD) to quantify variability due to training stochasticity.

\subsection{Experimental Setup and Notation}
For clarity in subsequent tables and analyses, we define shorthand notation for output formats, training sets, evaluation sets, and model names.

\subsubsection{Output formats}


We refer to the four extraction formats (Section~\ref{subsec:extraction}) using the following shorthand notation, and map each output to the evaluation categories in Table~\ref{tab:freq_real}:

\begin{itemize}
    \item \textbf{x per M} (Format~1): numeric seizures-per-month output, mapped to Purist/Pragmatic categories via Table~\ref{tab:freq_real}.
    \item \textbf{Categ.} (Format~2): four-way Pragmatic class output, evaluated under the Pragmatic scheme only.
    \item \textbf{Our label} (Format~3): Label Scheme output, deterministically converted to seizures per month and then mapped to Purist/Pragmatic categories.
    \item \textbf{Our CoT} (Format~4): CoT + label + evidence; scored identically to \textbf{Our label} using only the final predicted label (CoT and evidence spans are retained for interpretability).
\end{itemize}

\subsubsection{Training set}
All training datasets are constructed as instruction-formatted instances for supervised fine-tuning; each instance consists of a prompt template wrapping the clinic letter text and a format-specific target output. We define:

\begin{itemize}
    \item \textbf{R (x per M):}  
    1,481 instruction instances built from real clinic letters with clinician-provided numeric annotations. This dataset trains the model to output seizure frequency in the \textbf{x per M} label format.

    \item \textbf{R (Categ.):}  
    The same 1,481 real clinic letters as \textbf{R (x per M)}, converted into instruction instances that train the model to output \textbf{Categ.} labels. The categorical targets are derived from the original numeric annotations using the pragmatic frequency grouping in Table~\ref{tab:freq_real}. Compared with \textbf{R (x per M)}, this dataset differs in both the output format and the prompting instruction.

    \item \textbf{R+S (x per M):}  
    3,278 instruction instances consisting of the 1,481 instances from \textbf{R (x per M)} and 1,797 additional synthetic instances drawn from a different distribution. All instances train the model to output seizure frequency in the \textbf{x per M} label format.

    \item \textbf{R+S (Categ.):}  
    The same 3,278 instances as \textbf{R+S (x per M)}, converted into instruction instances that train the model to output \textbf{Categ.} labels using the pragmatic grouping in Table~\ref{tab:freq_real}.

    \item \textbf{CoT (1,548):}  
1,548 instruction instances sampled from the synthetic dataset, with the label distribution adjusted to better reflect the real training set \textbf{R (x per M)} and a dataset size comparable to it. This dataset is used to train the model to produce outputs in the \textbf{Our CoT} label format.

    \item \textbf{CoT (500 / 1,500 / 5,000 / 15,000):}  
    Randomly sampled synthetic instruction instances of the specified sizes, each training the model to output results in the \textbf{Our CoT} label format.
\end{itemize}

\subsubsection{Evaluation set}
We evaluate models on three complementary evaluation datasets:

\begin{itemize}
    \item \textbf{Real (300):}  
    A held-out set of 300 real KCH clinic letters that underwent double-checking by senior clinicians. The label distribution of this dataset closely matches that of the full real-world corpus and the training sets based on real clinic letters, and it serves as the primary in-distribution evaluation benchmark.

    \item \textbf{Real (150):}  
    A balanced subset of \textbf{Real (300)}, constructed to make category counts as even as possible across seizure-frequency classes. This dataset is designed to evaluate model robustness under a distribution that differs from the naturally imbalanced real-world setting.

    \item \textbf{Synthetic (1,166):}  
A synthetic evaluation set of 1,166 instruction instances was constructed from synthetic clinic letters. The instances were randomly sampled from the full pool of synthetic letters, while the frequency of the \emph{unknown} label was deliberately controlled to better reflect its distribution in \textbf{Real (300)}. This dataset is used to assess out-of-domain generalisation, testing whether models trained on real clinic letters can correctly interpret seizure-frequency expressions in fully synthetic but clinically realistic text. Importantly, this evaluation set has \emph{no overlap} with the synthetic training sets (1,548 and 1,791).
\end{itemize}

\subsubsection{Models}

For brevity, we use the following model abbreviations:

\begin{itemize}
    \item Meta-Llama-3.1-8B-Instruct (\textbf{Llama-3.1-8B})
    \item Qwen2.5-7B-Instruct (\textbf{Qwen2.5-7B})
    \item Qwen2.5-14B-Instruct (\textbf{Qwen2.5-14B})
    \item Ministral-8B-Instruct-2410 (\textbf{Ministral-8B})
    \item Gemma-3-4B-it (\textbf{Gemma-3-4B})
    \item Lingshu-7B (\textbf{Lingshu-7B})
    \item MedGemma-4B-it (\textbf{Medgemma-4B})
\end{itemize}

\section{Results}

\begin{table}[t]
\centering
\caption{Purist classification report for Llama-3.1-8B fine-tuned to output \textbf{Our CoT} and evaluated on the Real (300) dataset. The table highlights the severe class imbalance in fine-grained Purist categories, with several classes having very low support.}
\label{tab:purist_report}
\renewcommand{\arraystretch}{1.15}
\begin{tabular}{lrrrr}
\toprule
Class & Precision & Recall & F1-score & Support \\
\midrule
$<$1/6M& 0.2500 & 0.2000 & 0.2222 & 5 \\
= 1/6M & 0.0000 & 0.0000 & 0.0000 & 2 \\
(1/6M,1/M) & 0.5263 & 0.5263 & 0.5263 & 19 \\
= 1/M & 0.2000 & 0.3333 & 0.2500 & 6 \\
(1/M,1/W) & 0.5789 & 0.5789 & 0.5789 & 19 \\
= 1/W & 0.0000 & 0.0000 & 0.0000 & 4 \\
(1/W,1/D) & 0.7083 & 0.6538 & 0.6800 & 26 \\
$\ge$1/D & 0.6129 & 0.8261 & 0.7037 & 23 \\
UNK & 0.9156 & 0.8650 & 0.8896 & 163 \\
NS & 0.7429 & 0.7879 & 0.7647 & 33 \\
\midrule
accuracy & \multicolumn{3}{r}{0.7567} & 300 \\
macro avg & 0.4535 & 0.4771 & 0.4615 & 300 \\
weighted avg & 0.7657 & 0.7567 & 0.7590 & 300 \\
\bottomrule
\end{tabular}
\end{table}


\begin{table}[t]
\centering
\caption{Pragmatic classification report for Llama-3.1-8B fine-tuned to output \textbf{Our CoT} and evaluated on the Real (300) dataset. Compared with the Purist setting, category supports are more balanced, resulting in more stable per-class performance estimates.}
\label{tab:pragmatic_report}
\renewcommand{\arraystretch}{1.15}
\begin{tabular}{lrrrr}
\toprule
Class & Precision & Recall & F1-score & Support \\
\midrule
infrequent & 0.6471 & 0.6875 & 0.6667 & 32 \\
frequent & 0.8442 & 0.9028 & 0.8725 & 72 \\
UNK & 0.9156 & 0.8650 & 0.8896 & 163 \\

NS & 0.7429 & 0.7879 & 0.7647 & 33 \\
\midrule
accuracy & \multicolumn{3}{r}{0.8467} & 300 \\
macro avg & 0.7874 & 0.8108 & 0.7984 & 300 \\
weighted avg & 0.8508 & 0.8467 & 0.8480 & 300 \\
\bottomrule
\end{tabular}
\end{table}

\begin{sidewaystable*}
\renewcommand{\arraystretch}{1.2}   
\def\d{\hphantom{0}}
\caption{Performance of different LLMs fine-tuned on 1,797 synthetic letters and evaluated on \textbf{Real(300)}.}
\label{table: 1797 synthetic letters}%
\begin{tabular*}{\textheight}{@{\extracolsep\fill}llllllllllllll@{\extracolsep\fill}}%
\toprule
 & & \multicolumn{6}{c}{Purist} & \multicolumn{6}{c}{Pragmatic}\\
\cmidrule(lr){3-8} \cmidrule(lr){9-14}
model & Label & Micro  & Micro SD & Macro  & Macro SD & Weight & Weight SD & Micro  & Micro SD & Macro  & Macro SD & Weight & Weight SD\\
\midrule

Llama-3.1-8B &  Our COT & \underline{0.755} & 0.014 & \underline{0.466} & 0.03 & \underline{0.754} & 0.011 & \bf 0.828 & 0.015 & \bf 0.78 & 0.014 & \bf 0.829 & 0.014 \\
Llama-3.1-8B & Our label & \bf 0.758 & 0.02 & \bf 0.482 & 0.024 & \bf 0.758 & 0.018 & \underline{0.818} & 0.012 & \underline{0.768} & 0.012 & \underline{0.818} & 0.011 \\
Llama-3.1-8B & x per M & 0.691 & 0.009 & 0.392 & 0.014 & 0.679 & 0.009 & 0.764 & 0.01 & 0.677 & 0.015 & 0.769 & 0.008 \\
Llama-3.1-8B & Categ. & - & - & - & - & - & - & 0.767 & 0.013 & 0.581 & 0.009 & 0.729 & 0.011 \\

\hline

Qwen2.5-7B &  Our COT & \underline{0.756} & 0.01 & \bf{0.468} & 0.014 & \underline{0.755} & 0.011 & \underline{0.822} & 0.006 & \underline{0.776} & 0.008 & \underline{0.823} & 0.006 \\
Qwen2.5-7B & Our label & \bf{0.762} & 0.009 & \underline{0.465} & 0.007 & \bf{0.761} & 0.007 & \bf{0.832} & 0.009 & \bf{0.789} & 0.012 & \bf{0.833} & 0.008 \\
Qwen2.5-7B & x per & 0.715 & 0.023 & 0.388 & 0.021 & 0.695 & 0.022 & 0.785 & 0.025 & 0.673 & 0.033 & 0.782 & 0.025 \\
Qwen2.5-7B & Categ. & - & - & - & - & - & - & 0.768 & 0.008 & 0.588 & 0.009 & 0.731 & 0.008 \\
\hline

Ministral-8B &  Our COT & \bf{0.766} & 0.005 & \underline{0.475} & 0.014 & \bf{0.764} & 0.004 & \bf{0.828} & 0.002 & \underline{0.78} & 0.005 & \bf{0.829} & 0.003 \\
Ministral-8B & Our label& \underline{0.747} & 0.018 & \bf{0.518} & 0.027 & \underline{0.754} & 0.014 & \underline{0.818} & 0.017 & \bf{0.783} & 0.018 & \underline{0.82} & 0.017 \\
Ministral-8B & x per M & 0.718 & 0.009 & 0.399 & 0.005 & 0.699 & 0.009 & 0.775 & 0.01 & 0.667 & 0.019 & 0.771 & 0.012 \\
Ministral-8B & Categ. & - & - & - & - & - & - & 0.772 & 0.012 & 0.596 & 0.007 & 0.736 & 0.008 \\
\hline

Qwen2.5-14B &  Our COT & \bf{0.776} & 0.008 & \bf{0.497} & 0.031 & \bf{0.774} & 0.01 & \bf{0.832} & 0.008 & \bf{0.794} & 0.01 & \bf{0.833} & 0.008 \\
Qwen2.5-14B & Our label & \underline{0.745} & 0.018 & \underline{0.496} & 0.02 & \underline{0.751} & 0.014 & \underline{0.812} & 0.014 & \underline{0.763} & 0.01 & \underline{0.816} & 0.013 \\
Qwen2.5-14B & x per M & 0.713 & 0.014 & 0.391 & 0.009 & 0.689 & 0.013 & 0.763 & 0.015 & 0.633 & 0.019 & 0.756 & 0.016 \\
Qwen2.5-14B & Categ. & - & - & - & - & - & - & 0.74 & 0.011 & 0.567 & 0.008 & 0.708 & 0.01 \\
\hline

lingshu-7B &  Our COT & \underline{0.735} & 0.006 & \underline{0.439} & 0.005 & \underline{0.732} & 0.004 & \underline{0.802} & 0.008 & \underline{0.748} & 0.008 & \underline{0.802} & 0.008 \\
lingshu-7B & Our label & \bf{0.755} & 0.006 & \bf{0.464} & 0.02 & \bf{0.755} & 0.005 & \bf{0.822} & 0.007 & \bf{0.763} & 0.008 & \bf{0.822} & 0.007 \\
lingshu-7B & x per M & 0.681 & 0.019 & 0.391 & 0.015 & 0.67 & 0.015 & 0.759 & 0.012 & 0.685 & 0.017 & 0.766 & 0.011 \\
lingshu-7B & Categ. & - & - & - & - & - & - & 0.745 & 0.006 & 0.575 & 0.004 & 0.711 & 0.006 \\
\hline

Gemma-3-4B &  Our COT & \bf{0.75} & 0.009 & \bf{0.449} & 0.014 & \bf{0.744} & 0.009 & \bf{0.823} & 0.007 & \bf{0.766} & 0.009 & \bf{0.822} & 0.006 \\
Gemma-3-4B & Our label & \underline{0.717} & 0.024 & \underline{0.448} & 0.038 & \underline{0.72} & 0.02 & \underline{0.798} & 0.024 & \underline{0.748} & 0.028 & \underline{0.8} & 0.024 \\
Gemma-3-4B & x per M & 0.694 & 0.019 & 0.379 & 0.013 & 0.678 & 0.016 & 0.777 & 0.015 & 0.688 & 0.015 & 0.775 & 0.016 \\
Gemma-3-4B & Categ. & - & - & - & - & - & - & 0.781 & 0.006 & 0.591 & 0.006 & 0.743 & 0.006 \\
\hline

Medgemma-4b &  Our COT & \bf{0.75} & 0.012 & \bf{0.459} & 0.039 & \bf{0.745} & 0.011 & \bf{0.829} & 0.012 & \bf{0.784} & 0.018 & \bf{0.83} & 0.013 \\
Medgemma-4B& Our label & \underline{0.729} & 0.013 & \underline{0.455} & 0.022 & \underline{0.728} & 0.012 & \underline{0.797} & 0.009 & \underline{0.734} & 0.012 & \underline{0.796} & 0.01 \\
Medgemma-4B & x per M & 0.678 & 0.023 & 0.369 & 0.018 & 0.661 & 0.019 & 0.743 & 0.025 & 0.642 & 0.02 & 0.741 & 0.022 \\
Medgemma-4B & Categ. & - & - & - & - & - & - & 0.75 & 0.005 & 0.564 & 0.01 & 0.714 & 0.003 \\
\bottomrule
\end{tabular*}
\end{sidewaystable*}

\subsection{Purely Synthetic Letters}

Table~\ref{table: 1797 synthetic letters} reports the performance of seven open-weight LLMs fine-tuned \emph{only} on 1,797 synthetic clinic letters and evaluated on \textbf{Real (300)} under both the Purist and Pragmatic schemes. 
Following the evaluation protocol described in Section~\ref{subsec:evaluation}, we focus on micro F1 for primary comparisons, while macro and weighted averages are provided for completeness.

To contextualise metric behaviour under the fine-grained Purist scheme, Tables~\ref{tab:purist_report} and~\ref{tab:pragmatic_report} show representative per-class reports on Real (300) for Llama-3.1-8B trained with \textbf{Our CoT}. 
The Purist label distribution is highly imbalanced, with \texttt{UNK} comprising more than half of test instances (163/300) and several fine-grained categories having very low support. 
Consistent with this, per-class F1 estimates for rare Purist categories are highly variable, whereas the Pragmatic grouping yields more stable supports (32--163 per class) and correspondingly more stable per-class performance estimates.

Across model backbones, supervision with our structured outputs (\textbf{Our label} and \textbf{Our CoT}) achieved higher performance on Real (300) than simplified targets (\textbf{x per M} and \textbf{Categ.}), under both Purist and Pragmatic evaluation.
In addition, \textbf{Our CoT} provides explicit rationales and evidence spans, supporting qualitative review of model decisions.

\begin{sidewaystable*}
\renewcommand{\arraystretch}{1.05}   
\def\d{\hphantom{0}}
\caption{Micro F1 results of different models across different training sets and evaluation datasets.\label{tab:different-train-different-eval}}%
\begin{tabular*}{\textheight}{@{\extracolsep\fill}llllllllllllll@{\extracolsep\fill}}%
\toprule
 & & \multicolumn{4}{c}{Real(300)} & \multicolumn{4}{c}{Real(150)}& \multicolumn{4}{c}{Synthetic(1166)}\\
\cmidrule(lr){3-6} \cmidrule(lr){7-10} \cmidrule(lr){11-14}
model & Training set & Purist & SD & Pragmatic  &  SD & Purist & SD & Pragmatic & SD & Purist & SD & Pragmatic & SD\\
\midrule

Llama‐3.1‐8B & COT(1548) & 0.755 & 0.01 & 0.828 & 0.01 & 0.625 & 0.026 & \underline{0.773} & 0.02 & \underline{0.738} & 0.011 & 0.81 & 0.008 \\
Llama‐3.1‐8B & R(x per M) & \underline{0.782} & 0.015 & 0.839 & 0.01 & \bf 0.66 & 0.014 & 0.77 & 0.025 & 0.513 & 0.006 & 0.682 & 0.012\\
Llama‐3.1‐8B & R(Categ.) & - & - & \underline{0.846} & 0.008 & - & - & \bf 0.783 & 0.009 & - & - & 0.673 & 0.018 \\
Llama‐3.1‐8B & R+S(x per M) & \bf{0.785} & 0.014 & \bf 0.857 & 0.014 & \underline{0.633} & 0.016 & 0.74 & 0.011 &  \bf  0.796 & 0.009 & \underline{0.872} & 0.011  \\
Llama‐3.1‐8B & R+S(Categ.) & - & - & 0.845 & 0.009 & - & - & 0.75 & 0.012 &- & - & \bf 0.886 & 0.018\\
\hline

Qwen‐2.5‐7B & COT(1548) & \underline{0.759} & 0.003 & 0.828 & 0.006 & \bf 0.65 & 0.016 & 0.777 & 0.012 & \underline{0.702} & 0.006 & 0.777 & 0.01  \\
Qwen‐2.5‐7B & R(x per M) & \bf{0.765} & 0.017 & 0.833 & 0.012 & 0.608 & 0.01 & 0.727 & 0.024 & 0.503 & 0.011 & 0.624 & 0.028  \\
Qwen‐2.5‐7B & R(Categ.) & - & - & \bf{0.861} & 0.007 & - & - & \bf{0.818} & 0.006 & - & - & 0.686 & 0.016\\
Qwen‐2.5‐7B & R+S(x per M) & \underline{0.759} & 0.014 & 0.827 & 0.013 & \underline{0.617} & 0.019 & 0.717 & 0.025 & \bf 0.784 & 0.011 & \underline{0.855} & 0.011 \\
Qwen‐2.5‐7B & R+S(Categ.) & - & - & \underline{0.857} & 0.007 & - & - & \underline{0.788} & 0.008 &  - & - & \bf 0.881 & 0.009 \\

\hline

Ministral‐8B & COT(1548) & 0.763 & 0.014 & 0.827 & 0.007 & 0.645 & 0.008 & 0.772 & 0.015 & \underline{0.808} & 0.006 & 0.869 & 0.005 \\
Ministral‐8B & R(x per M) & \bf 0.801 & 0.007 & 0.846 & 0.003 & \bf 0.682 & 0.013 & 0.758 & 0.016 & 0.552 & 0.006 & 0.676 & 0.005\\
Ministral‐8B & R(Categ.) & - & - & \bf 0.863 & 0.011 & - & - & \bf{0.787} & 0.009 & - & - & 0.70 & 0.013 \\
Ministral‐8B & R+S(x per M) & \underline{0.79} & 0.008 & 0.839 & 0.01 & \underline{0.653} & 0.012 & 0.735 & 0.008 & \bf{0.812} & 0.013 & \underline{0.883} & 0.014\\
Ministral‐8B & R+S(Categ.) & - & - & \underline{0.85} & 0.011 & - & - & \underline{0.775} & 0.015 & - & - & \bf{0.908} & 0.004 \\

\hline

Qwen‐2.5‐14B & COT(1548) & 0.757 & 0.012 & 0.816 & 0.011 & \bf{0.652} & 0.014 & \underline{0.76} & 0.014 & \underline{0.782} & 0.003 & 0.848 & 0.003 \\
Qwen‐2.5‐14B & R(x per M) & \underline{0.763} & 0.008 & 0.811 & 0.009 & 0.612 & 0.006 & 0.708 & 0.016 & 0.562 & 0.004 & 0.666 & 0.01 \\
Qwen‐2.5‐14B & R(Categ.) & - & - & \bf{0.831} & 0.003 & - & - & 0.733 & 0.008 &  - & - & 0.694 & 0.005 \\
Qwen‐2.5‐14B & R+S(x per M) & \bf{0.77} & 0.005 & 0.818 & 0.004 & \underline{0.623} & 0.013 & 0.697 & 0.012 & \bf{0.817} & 0.009 & \bf{0.889} & 0.011 \\
Qwen‐2.5‐14B & R+S(Categ.) & - & - & \underline{0.83} & 0.005 & - & - & \bf{0.762} & 0.006 & - & - & \underline{0.870} & 0.02 \\

\hline

Lingshu‐7B & COT(1548) & 0.747 & 0.003 & 0.814 & 0.007 & \bf{0.64} & 0.021 & 0.76 & 0.011 & \underline{0.717} & 0.001 & 0.798 & 0.004\\
Lingshu‐7B & R(x per M) & \underline{0.768} & 0.007 & 0.827 & 0.008 & \underline{0.637} & 0.016 & 0.745 & 0.006 & 0.520 & 0.01 & 0.645 & 0.022\\
Lingshu‐7B & R(Categ.) & - & - & \underline{0.833} & 0.006 & - & - & \bf{0.777} & 0.004 & - & - & 0.684 & 0.004 \\
Lingshu‐7B & R+S(x per M) & \bf{0.775} & 0.01 & 0.832 & 0.01 & 0.623 & 0.021 & 0.728 & 0.006 & \bf{0.758} & 0.025 & \underline{0.831} & 0.017\\
Lingshu‐7B & R+S(Categ.) & - & - & \bf{0.848} & 0.006 & - & - & \underline{0.773} & 0.014 & - & - & \bf{0.881} & 0.012 \\

\hline

Gemma‐3‐4B & COT(1548) & 0.738 & 0.006 & 0.802 & 0.01 & \underline{0.598} & 0.011 & 0.73 & 0.013 & \underline{0.707} & 0.001 & 0.786 & 0.005\\
Gemma‐3‐4B & R(x per M) & \underline{0.741} & 0.011 & 0.808 & 0.009 & 0.588 & 0.015 & 0.715 & 0.015 & 0.451 & 0.033 & 0.608 & 0.028 \\
Gemma‐3‐4B & R(Categ.) & - & - & \underline{0.842} & 0.007 & - & - & \bf{0.767} & 0.012 & - & - & 0.682 & 0.004\\
Gemma‐3‐4B & R+S(x per M) & \bf{0.749} & 0.011 & 0.817 & 0.012 & \bf{0.618} & 0.003 & 0.723 & 0.014 & \bf{0.762} & 0.024 & \underline{0.844} & 0.022  \\
Gemma‐3‐4B & R+S(Categ.) & - & - & \bf{0.847} & 0.004 & - & - & \underline{0.753} & 0.014 & - & - & \bf{0.896} & 0.005 \\

\hline

MedGemma‐4B & COT(1548) & 0.741 & 0.007 & 0.819 & 0.007 & \bf 0.58 & 0.016 & \underline{0.732} & 0.026 & \underline{0.693} & 0.007 & 0.770 & 0.007  \\
MedGemma‐4B & R(x per M) & \underline{0.743} & 0.02 & 0.817 & 0.016 & 0.55 & 0.027 & 0.653 & 0.037 & 0.446 & 0.028 & 0.543 & 0.039 \\
MedGemma‐4B & R(Categ.) & - & - & \bf{0.83} & 0.014 & - & - & 0.73 & 0.026 & - & - & 0.662 & 0.011  \\
MedGemma‐4B & R+S(x per M) & \bf{0.752} & 0.008 & \underline{0.833} & 0.007 & \underline{0.57} & 0.026 & 0.688 & 0.031 & \bf{0.752} & 0.034 & \underline{0.832} & 0.026  \\
MedGemma‐4B & R+S(Categ.) & - & - & 0.828 & 0.01 & - & - & \bf{0.747} & 0.011 & - & - & \bf{0.884} & 0.015 \\
\end{tabular*}
\end{sidewaystable*}

\subsection{Synthetic versus Real Training Data}
Table~\ref{tab:different-train-different-eval} reports Micro F1 scores for all models across different training sets and evaluation datasets, under both the Purist and Pragmatic evaluation schemes. Results are grouped by evaluation dataset (\textbf{Real(300)}, \textbf{Real(150)}, and \textbf{Synthetic(1,166)}) to highlight how training data composition affects generalisation performance under different distributional settings.
For each model, the best-performing result is highlighted in bold, and the second-best result is underlined for emphasis.

\begin{table*}[t]
\renewcommand{\arraystretch}{1.25}
\caption{Scaling behaviour with increasing amounts of synthetic training data.}
\label{tab:Scaling}
\centering
\begin{tabular}{llllllllll}
\toprule
 & & \multicolumn{4}{c}{Real(300)} & \multicolumn{4}{c}{Real(150)}\\
\cmidrule(lr){3-6} \cmidrule(lr){7-10}
model & Method & Purist & SD & Pragmatic  &  SD & Purist & SD & Pragmatic & SD\\
\midrule

Qwen‐2.5‐14B & COT(500) & 0.743 & 0.015 & 0.814 & 0.007 & 0.662 & 0.015 & \bf{0.8} & 0.012 \\
Qwen‐2.5‐14B & COT(1500) & 0.762 & 0.016 & 0.827 & 0.016 & 0.665 & 0.01 & 0.795 & 0.01 \\
Qwen‐2.5‐14B & COT(5000) & \underline{0.77} & 0.007 & \underline{0.832} & 0.007 & \underline{0.668} & 0.021 & \underline{0.797} & 0.018 \\
Qwen‐2.5‐14B & COT(15000) & \bf 0.788 & 0.01 & \bf 0.847 & 0.006 & \bf{0.695} & 0.008 & \bf{0.8} & 0.02 \\
Qwen‐2.5‐14B & R(x per M) & 0.763 & 0.008 & 0.811 & 0.009 & 0.612 & 0.006 & 0.708 & 0.016 \\
Qwen‐2.5‐14B & R(Categ.) & - & - & 0.831 & 0.003 & - & - & 0.733 & 0.008 \\
Qwen‐2.5‐14B & R+S(x per M) & \underline{0.77} & 0.005 & 0.818 & 0.004 & 0.623 & 0.013 & 0.697 & 0.012 \\
Qwen‐2.5‐14B & R+S(Categ.) & - & - & 0.83 & 0.005 & - & - & 0.762 & 0.006 \\
\hline
MedGemma‐4B & COT(500) & 0.708 & 0.006 & 0.792 & 0.005 & 0.583 & 0.017 & 0.75 & 0.027 \\
MedGemma‐4B & COT(1500) & 0.751 & 0.01 & 0.827 & 0.005 & 0.627 & 0.022 & 0.773 & 0.018 \\
MedGemma‐4B & COT(5000) & \underline{0.777} & 0.011 & \underline{0.844} & 0.01 & \underline{0.695} & 0.021 & \underline{0.835} & 0.01 \\
MedGemma‐4B & COT(15000) & \bf{0.787} & 0.014 & \bf{0.858} & 0.01 & \bf{0.702} & 0.014 & \bf{0.84} & 0.009 \\
MedGemma‐4B & R(x per M) & 0.743 & 0.02 & 0.817 & 0.016 & 0.55 & 0.027 & 0.653 & 0.037 \\
MedGemma‐4B & R(Categ.) & - & - & 0.83 & 0.014 & - & - & 0.73 & 0.026 \\
MedGemma‐4B & R+S(x per M) & 0.752 & 0.008 & 0.833 & 0.007 & 0.57 & 0.026 & 0.688 & 0.031 \\
MedGemma‐4B & R+S(Categ.) & - & - & 0.828 & 0.01 & - & - & 0.747 & 0.011 \\
\hline
\bottomrule
\end{tabular}
\end{table*}

\begin{table}[t]
\renewcommand{\arraystretch}{1.25}
\caption{Performance when trained on 15,000 synthetic examples (COT format) and evaluated on the real training set.}
\label{tab:eval_trainset_synth15000}
\centering
\begin{tabular}{lllll}
\toprule
model  & Purist & SD & Pragmatic  &  SD \\
\midrule
Qwen-2.5-14B &  0.779 & 0.006 & 0.835 & 0.005 \\
MedGemma‐4B & 0.762 & 0.002 & 0.819 & 0.005\\ 

\hline
\bottomrule
\end{tabular}
\end{table}

\subsection{Scaling Behaviour with Synthetic Letters}

Table~\ref{tab:Scaling} examines how extraction performance changes as the amount of synthetic training data increases, while keeping the model architecture and output format fixed. We focus on two representative models with different parameter scales and domain specialisation, \textbf{Qwen-2.5-14B} and \textbf{MedGemma-4B}, and vary the size of the synthetic CoT training set from 500 to 15,000 letters.
For each model, the best-performing result is highlighted in bold, and the second-best result is underlined for emphasis.
\subsection{Evaluation on Synthetic Letters}
\label{sec:synthetic_letter_eval}

To assess the realism of the synthetic letters, two neurologists independently evaluated a blinded set of clinic letters, each of which was either real (clinician-written) or synthetic. For each letter, raters selected one of three options: (i) written by a doctor, (ii) synthetic but doctor-like, or (iii) synthetic and not doctor-like. Each neurologist assessed the same 60 letters.

Table~\ref{tab:turing_test} summarises the results by rater. Real letters were consistently identified as doctor-written (100\% recall for both raters). For synthetic letters, Neurologist~1 rated 56.7\% synthetic letters as doctor-like (options~i and ii), with 3.3\% strictly judged as doctor-written (option~i). Neurologist~2 rated 86.7\% synthetic letters as doctor-like, with 23.3\% judged as doctor-written.

Qualitative feedback indicated that the synthetic letters were clinically coherent and generally convincing. 
Nevertheless, both neurologists identified recurring stylistic artefacts, including:
(i) an overly comprehensive, insurance-oriented (US-like) documentation style; 
(ii) inclusion of investigations or examination details uncommon in routine UK epilepsy follow-up (e.g., broad blood testing or detailed cardiovascular/respiratory examination); 
(iii) frequent use of phrasing such as ``per''; 
(iv) occasionally implausibly short follow-up intervals; and 
(v) an overly structured narrative flow compared with the more variable dictation style typical of real correspondence.

Overall, although the synthetic letters still differ in some stylistic aspects from routine UK NHS correspondence, these discrepancies do not appear to undermine their utility for model training. In particular, despite the identified locale-specific artefacts, the models trained on this synthetic corpus were still able to learn the seizure-frequency extraction task effectively, suggesting that performance is robust to moderate variation in writing style and documentation conventions.

\begin{table}[t]
\centering
\small
\setlength{\tabcolsep}{5pt}
\begin{tabular}{lcccc}
\toprule
\textbf{Rater} &
\textbf{Acc.} &
\textbf{Real rec.} &
\textbf{Dr-written} &
\textbf{Dr-like} \\
\midrule
1 & 98.3\% & 100.0\% & 3.3\%  & 56.7\% \\
2 & 88.3\% & 100.0\% & 23.3\% & 86.7\% \\
\bottomrule
\end{tabular}
\caption{Blinded clinician realism assessment of real vs.\ synthetic letters (60 letters; same set rated by both neurologists). Acc.: binary real/synthetic accuracy (option~i vs.\ options~ii--iii). Real rec.: recall on real letters. Dr-written: synthetic letters rated as doctor-written (option~i). Dr-like: synthetic letters rated as doctor-like (options~i--ii).}
\label{tab:turing_test}
\end{table}

\section{Discussion}

\subsection{Principal findings}
In this study, we evaluated open-weight large language models (LLMs) for extracting seizure frequency from UK NHS-style epilepsy clinic letters under realistic NHS deployment constraints. Our results highlight four main findings.

First, seizure-frequency evaluation on real-world clinic letters is strongly affected by class imbalance, particularly under the fine-grained Purist scheme. In the Real (300) test set, more than half of letters were labelled as \texttt{UNK} (163/300), whereas several Purist categories had extremely low support (e.g., \texttt{=1/6M} with only two instances). As a result, aggregate metrics that weight rare classes equally can be unstable, and overall performance is best summarised using Micro F1 alongside per-class results for clinical error interpretation.

Second, output representation substantially influences trainability and downstream performance. When training exclusively on 1,797 synthetic letters, structured outputs (Our label / Our CoT) consistently outperformed simpler targets (numeric \emph{x per month} and four-way Pragmatic categorisation) on Real (300) across model families. This suggests that explicitly modelling the linguistic structure of seizure-frequency statements is advantageous for learning robust mappings from narrative text to standardised frequency labels.

Third, training-data composition markedly affected robustness under distribution shift. Models trained only on real clinic letters performed well in-distribution (Real (300)) but exhibited large performance drops when evaluated on synthetic letters, whereas models trained with synthetic augmentation (R+S) or trained purely on synthetic CoT data retained substantially higher performance on the Synthetic (1,166) evaluation set. In addition, evaluation on the balanced Real (150) subset reduced performance for many settings, reinforcing that robustness should be assessed beyond naturally imbalanced test distributions.

Finally, scaling synthetic CoT training data improved performance for both a larger general-purpose model (Qwen2.5-14B) and a smaller medically oriented model (MedGemma-4B). With 15,000 synthetic CoT training letters, both models achieved their best overall performance on Real (300) and Real (150), and in several settings exceeded models trained on real letters alone. Notably, this scaling behaviour indicates that high-quality synthetic supervision can partially compensate for limited access to real patient text, while preserving privacy and enabling reproducible training.

\subsection{Interpretation and comparison with prior work}
Prior work has shown that extracting epilepsy-related attributes from unstructured clinic letters is feasible with open-weight LLMs, but performance can be sensitive to prompt design and task complexity, and remains below specialist human annotation for nuanced categories \citep{Fang2025Epilepsia,holgate-etal-2025-fine}. Our work extends this literature in two key ways.

First, seizure frequency is intrinsically more temporally complex than many discrete epilepsy attributes (e.g., medication lists), because documentation often includes ranges, vague quantifiers, changing time windows, and ambiguous references. A direct numeric target (\emph{x per month}) compresses this uncertainty into a single value, which may increase the implicit reasoning burden during decoding and make errors harder to audit. In contrast, our Label Scheme (and its CoT extension) preserves clinically salient structure (rates, ranges, seizure-free durations, and cluster patterns) while remaining machine-parsable and deterministically mappable to standardised units. The consistent gains observed for structured labels over numeric and coarse categorical targets support the hypothesis that an output format aligned with clinical language can improve learnability and stability.

Second, our results suggest that robustness is not guaranteed by training on real letters alone. The large generalisation gap observed when real-trained models were evaluated on synthetic letters indicates that some models may rely on distribution-specific cues in the real training set rather than learning generalisable seizure-frequency reasoning. Synthetic augmentation (R+S) improved out-of-domain performance, but requires access to extra letters for training. 
Importantly, purely synthetic CoT training achieved strong performance on real held-out letters while maintaining robustness across evaluation sets, indicating that training with synthetic data alone can be viable when the synthetic data are diverse, clinically realistic, and paired with high-quality supervision signals.

\subsection{Value of CoT and evidence grounding for clinical interpretability}
Beyond accuracy, clinical adoption requires transparency and auditability. Our CoT format augments the structured label with intermediate reasoning and explicit evidence spans copied from the source letter. This design enables rapid clinical verification: reviewers can check whether the predicted frequency is supported by the cited text, and can distinguish failures due to evidence localisation from failures due to temporal normalisation. Although CoT supervision did not uniformly dominate structured labels in every model setting, it provided comparable performance while offering an interpretable trace, supporting its role as a practical compromise between performance and trustworthiness in clinical NLP.

\subsection{Synthetic data scaling and data diversity}
The scaling experiments demonstrate that increasing synthetic CoT training size improves performance, with diminishing incremental gains beyond several thousand examples. A plausible explanation is limited diversity in the synthetic-generation seed set: synthetic letters were derived from a small number of base templates combined with templated seizure-frequency descriptions. Under such constraints, enlarging the dataset can increase coverage of surface forms and numeric values, but may eventually saturate stylistic and contextual diversity. 
Nevertheless, because synthetic data generation is comparatively low-cost, these results are encouraging. Further improvements may be achievable by increasing the diversity of base letters, expanding clinical contexts beyond seizure-focused paragraphs (e.g., comorbidities, investigations, and medication changes), and introducing controlled variation in documentation style, thereby better capturing cross-clinician and cross-site heterogeneity.

Table~\ref{tab:eval_trainset_synth15000} reports the performance of models trained on 15,000 synthetic letters when evaluated on the real-letter training set. The results show that the F1 scores under both the purist and pragmatic settings are very close to those obtained on the real-letter test set (300). This demonstrates that training on our synthetic data generalises well to real-world letters and that the proposed synthetic-data training strategy is both reliable and effective.


\subsection{Strengths and limitations}
This study has several strengths. We evaluate multiple open-weight models under consistent training and decoding settings, use a double-checked real-world gold-standard test set, and propose a structured label representation that supports both quantitative normalisation and clinical interpretability. The synthetic data framework provides a reproducible, privacy-preserving pathway to scale supervision without exposing patient text.

Several limitations warrant emphasis. First, real clinic letters were drawn from a single UK centre and a fixed historical period; external validation across hospitals, regions, and documentation systems is required before broad deployment. Second, the real-letter annotations were available only in numeric form, preventing joint training of structured-label formats on real data; future work could explore efficient annotation or semi-automated conversion strategies for real letters. Third, the synthetic corpus was produced via a constrained template-based pipeline, which may not fully capture the diversity of real clinical narratives; this may partly explain diminishing returns at larger synthetic scales. Finally, although CoT and evidence spans provide interpretability, we did not conduct a prospective clinician user study to quantify time savings, trust, or error detection during routine review.

\subsection{Future work}
Future research should prioritise (i) extending information extraction beyond seizure frequency to additional clinically relevant attributes; (ii) expanding synthetic data generation to increase the diversity of clinical contexts and documentation styles; (iii) investigating strategies for continued training and adaptation of models trained on synthetic data using real-world clinic letters, in settings where CoT rationales and evidence-span annotations are unavailable; and (iv) human-in-the-loop evaluation to assess how evidence-grounded outputs affect clinician workload, confidence, and safety. In addition, integrating uncertainty-aware outputs (e.g., ranges or calibrated confidence estimates) may better reflect the ambiguity inherent in seizure-frequency documentation.

\subsection{Conclusions}
In summary, we show that seizure-frequency extraction from unstructured epilepsy clinic letters can be improved by using structured output representations and scalable, privacy-preserving synthetic supervision. Synthetic CoT training yields strong performance on real held-out letters, improves robustness under distribution shift, and provides interpretable evidence-grounded outputs suitable for clinical audit. These results support the feasibility of reproducible, open-weight LLM pipelines for epilepsy research workflows, while underscoring the need for external validation and continued clinician oversight for clinical use.

\bibliographystyle{model1-num-names}

\bibliography{cas-refs}





\end{document}